\documentclass[12pt,nofootinbib]{article}

\usepackage{graphicx}
\usepackage{amssymb}
\usepackage{amsmath}
\usepackage{color}
\usepackage[colorlinks=true,linkcolor=blue,citecolor=blue]{hyperref}

\begin{document}
\newcommand{\rem}[1]{{\bf [#1]}} \newcommand{\gsim}{ \mathop{}_
{\textstyle \sim}^{\textstyle >} } \newcommand{\lsim}{ \mathop{}_
{\textstyle \sim}^{\textstyle <} } \newcommand{\vev}[1]{ \left\langle
{#1} \right\rangle } \newcommand{\bear}{\begin{array}} \newcommand
{\eear}{\end{array}} \newcommand{\bea}{\begin{eqnarray}}
\newcommand{\eea}{\end{eqnarray}} \newcommand{\beq}{\begin{equation}}
\newcommand{\eeq}{\end{equation}} \newcommand{\bef}{\begin{figure}}
\newcommand {\eef}{\end{figure}} \newcommand{\bec}{\begin{center}}
\newcommand {\eec}{\end{center}} \newcommand{\non}{\nonumber}
\newcommand {\eqn}[1]{\beq {#1}\eeq} \newcommand{\la}{\left\langle}
\newcommand{\ra}{\right\rangle} \newcommand{\ds}{\displaystyle}
\newcommand{\red}{\textcolor{red}} 
\def\SEC#1{Sec.~\ref{#1}} \def\FIG#1{Fig.~\ref{#1}}
\def\EQ#1{Eq.~(\ref{#1})} \def\EQS#1{Eqs.~(\ref{#1})} \def\lrf#1#2{
\left(\frac{#1}{#2}\right)} \def\lrfp#1#2#3{ \left(\frac{#1}{#2}
\right)^{#3}} \def\GEV#1{10^{#1}{\rm\,GeV}}
\def\MEV#1{10^{#1}{\rm\,MeV}} \def\KEV#1{10^{#1}{\rm\,keV}}
\def\REF#1{(\ref{#1})} \def\lrf#1#2{ \left(\frac{#1}{#2}\right)}
\def\lrfp#1#2#3{ \left(\frac{#1}{#2} \right)^{#3}} \def\OG#1{ {\cal
O}(#1){\rm\,GeV}}

\begin{titlepage}

\begin{flushright}
UCB-PTH-15/18 \\
IPMU16-0003
\end{flushright}

\begin{center}

\vskip 1.2cm

{\usefont{T1}{ppl}{m}{n}
{\Large Cosmology with a Heavy Polonyi Field
} 
}

\vskip 1.2cm

{\usefont{T1}{bch}{m}{n}
Keisuke Harigaya$^{a,b}$,
Taku Hayakawa$^c$, 
Masahiro Kawasaki$^{c,d}$
and
Masaki Yamada$^{c,d}$}

\vskip 0.4cm

{\it$^a$Department of Physics, University of California,
Berkeley, California 94720, USA}\\
{\it$^b$Theoretical Physics Group, Lawrence Berkeley National Laboratory,
Berkeley, California 94720, USA}\\
{\it$^c$Institute for Cosmic Ray Research, The University of Tokyo,
5-1-5 Kashiwanoha, Kashiwa, Chiba 277-8582, Japan}\\
{\it$^d$Kavli IPMU (WPI), UTIAS, The University of Tokyo, 5-1-5 Kashiwanoha, 
Kashiwa, 277-8583, Japan}\\

\date{\today}

\vskip 1.2cm

\begin{abstract}
We consider a cosmologically consistent scenario with a heavy Polonyi field.
The Polonyi field with a mass of ${\cal O}(100){\rm\,TeV}$ decays
before the Big-Bang Nucleosynthesis (BBN)
and avoids the severe constraint from the BBN.
However,
the abundance of the Lightest Supersymmetric Particle (LSP) 
produced from the decay often exceeds the observed dark matter density.
In our scenario,
the dark matter density is obtained by the LSP abundance with an aid of entropy production,
and baryon asymmetry is generated by the Affleck-Dine mechanism.
We show that the observed baryon-to-dark matter ratio of ${\cal O}(0.1\mathchar`-1)$
is naturally explained in sequestering models with a QCD axion.
\end{abstract}

\end{center}
\end{titlepage}

\baselineskip 6mm

\section{Introduction}
\label{sec:intro}

Cosmological observations have revealed 
the existence of the dark matter (DM) and the baryon asymmetry,
while their origins remain unknown for a long time.
The cosmic densities of the DM and baryonic components have been
precisely measured by the observation of the Cosmic Microwave Background (CMB) radiation,
and the observed baryon-to-DM ratio is 
$\Omega_{B}/\Omega_{\rm DM}=0.18$~\cite{Agashe:2014kda}.
We need a cosmologically consistent scenario
explaining both the baryon asymmetry and the DM abundance.

Since those origins cannot be explained in the framework of the Standard Model (SM),
there should be underlying physics beyond the SM.
Supersymmetry (SUSY)~\cite{Nilles:1983ge} is one of the most interesting models
since it achieves the unification of the SM gauge couplings
and can relax the hierarchy problem.
In addition,
supersymmetric models with a conserved $R$ parity predict the stability of
the Lightest Supersymmetric Particle (LSP),
which becomes a good candidate for DM.
Moreover,
SUSY extensions of the SM contain a lot of flat directions with $B-L$ charges~\cite{Gherghetta:1995dv},
which can produce $B-L$ asymmetry~\cite{Affleck:1984fy}.
In the early universe,
one of the flat directions,
which we call the Affleck-Dine field,
may receive an angular kick from SUSY breaking and $R$ symmetry breaking effects
and rotates in its complex plane,
which corresponds to the generation of the $B-L$ asymmetry.
The $B-L$ asymmetry is converted into the baryon asymmetry 
through the sphaleron process~\cite{Kuzmin:1985mm,Khlebnikov:1988sr}.
This mechanism,
known as ``the Affleck-Dine (AD) mechanism"~\cite{Affleck:1984fy,Dine:1995kz},
can produce baryon number more effectively than most baryogenesis scenarios.
For these advantages,
we focus on supersymmetric theories in this paper.

The SUSY must be spontaneously broken
since superparticles have not been discovered yet.
The simplest SUSY breaking model
is the Polonyi model~\cite{Polonyi:1} in which an $F$-term of an elementary singlet field $Z$
breaks the SUSY in the hidden sector.\footnote{In the Polonyi model, 
the SUSY breaking scale can be easily obtained by dynamical transmutation, 
by assigning a vanishing $R$ charge to the Polonyi field and breaking the $R$ symmetry by gaugino condensation.
}
This model is attractive because of its simplicity.
However,
such a singlet field $Z$, called the Polonyi field,
may cause a cosmological difficulty~\cite{Coughlan:1983ci}.
Since the Polonyi potential has no symmetry enhanced points,
the minimum of the potential during the inflation is deviated from the true vacuum.
After inflation,
the Polonyi field begins to oscillate around the true minimum 
with an amplitude of the order of the Planck scale.
The energy density of its coherent oscillation immediately dominates the universe
after the inflaton decays. 
Since the Polonyi field very weakly interacts with particles in the observable sector,
its late-time decay upsets the standard scenario of the Big-Bang Nucleosynthesis (BBN),
which is referred to as ``the Polonyi problem".

There are mainly two ways of solving the problem.
One possible way is to dilute the Polonyi density by some mechanisms,
for example, by thermal inflation~\cite{Lyth:1995ka}.
In this case,
however,
pre-existing baryon asymmetry is also diluted,
and it is known to be difficult to produce a sufficient amount of baryon number beforehand.
Therefore,
baryon asymmetry should be produced after the dilution,
and we need an intricate model proposed 
in the context of the thermal inflation~\cite{Stewart:1996ai,Jeong:2004hy,Kawasaki:2006py}.
The other simple solution is to make the Polonyi field heavy enough to decay before the BBN.
Even in this case,
we need some dilution
since the abundance of LSPs produced from the Polonyi decay
exceeds the observed DM abundance unless pair annihilation is very effective~\cite{Moroi:1999zb,Moroi:2013sla}.
When such dilution is needed,
the AD mechanism is the leading candidate for baryogenesis
since most baryogensis scenarios including the thermal leptogenesis~\cite{Fukugita:1986hr} 
cannot produce a sufficient amount of baryon asymmetry beforehand.

In this paper,
we consider the case where the Polonyi field is heavy enough to decay before the BBN,
i.e.,
$m_Z\sim{\cal O}(100){\rm TeV}$,
where $m_Z$ is a mass of the Polonyi field.
We construct a cosmologically consistent scenario in the presence of the heavy Polonyi field.
In our scenario,
the baryon asymmetry is created by dynamics of the AD field
which takes the vacuum expectation value (VEV)
of the order of the Planck scale in the early universe.
Both the Polonyi density and the baryon asymmetry are diluted by late-time entropy production, 
for example, by the thermal inflation.
After the entropy production,
the Polonyi field decays into LSPs
which explain the observed DM abundance.
We show that the baryon-to-DM ratio is simply determined by the LSP mass
and the branching fraction of the decay of the Polonyi field into superparticles.
The correct baryon-to-DM ratio is obtained 
when the LSP mass is of ${\cal O}(1){\rm\,TeV}$ 
and the branching fraction of the decay of the Polonyi into superparticles is of ${\cal O}(10^{-3})$.

In order to realize the branching fraction of ${\cal O}(10^{-3})$,
we consider a specific model satisfying the following conditions.
Firstly,
we assume that
the SUSY breaking sector is sequestered from the visible sector in superspace Lagrangian,
which is called ``the sequestering model"~\cite{Inoue:1991rk,Randall:1998uk}.
The squarks/sleptons and gauginos have vanishing masses at the tree level
and acquire loop-suppressed masses through quantum corrections, 
such as anomaly-mediated SUSY breaking effects~\cite{Randall:1998uk,Giudice:1998xp}
(see also Refs.~\cite{Bagger:1999rd,Harigaya:2014sfa})
or one loop corrections from Planck-suppressed interactions~\cite{Randall:1998uk,Antoniadis:1997ic}.
Sfermions and gauginos are lighter than the gravitino with a mass of ${\cal O}(100)${\rm\,TeV}.
In addition,
we introduce
a (pseudo-)Nambu-Goldstone boson (NGB).
The NGB can be identified with the QCD axion~\cite{Peccei:1977hh,Peccei:1977ur,Weinberg:1977ma,Wilczek:1977pj},
which solves the strong CP problem~\cite{'tHooft:1976up,Jackiw:1976pf,Callan:1976je}.
In this model,
the Polonyi field mainly decays into NGBs,
but their contribution to the DM abundance is negligible
since the NGB is much lighter than the LSP (as is the case with the QCD axion) or decays into SM particles.
On the other hand,
the decay of the Polonyi into superparticles is suppressed for the sequestered potential,
and the branching ratio is of ${\cal O}(10^{-3})$.
The DM abundance is determined by the LSP abundance produced through the Polonyi decay.

This paper is organized as follows.
In Sec.~\ref{sec:Polonyi},
we briefly review the Polonyi problem
and explain why the entropy production is needed.
We explain the AD mechanism in Sec.~\ref{sec:Affleck-Dine}.
In Sec.~\ref{sec:B-DM},
we show that the baryon-to-DM ratio is simply determined by the LSP mass and
the branching fraction.
We also introduce the sequestering model as a candidate to realize our scenario.
The final section is devoted to summary and discussions.

\section{The Polonyi Problem}
\label{sec:Polonyi}

In this section,
we explain the cosmological problem
of the Polonyi model.

Let us
briefly introduce the Polonyi model~\cite{Polonyi:1}.
In this model,
the only ingredient in the SUSY breaking sector is 
an elementary field $Z$ which is singlet under any symmetry.
We call it as the Polonyi field.
The superpotential in the hidden sector is given by\footnote{Hereafter,
we use the same letter $Z$ to denote the scalar component of the supermultiplet.
}
\begin{equation}
	W_{\rm hid}=\mu^2M_{\rm pl}\left(1+c\frac{Z}{M_{\rm pl}}+\cdots\right),
\end{equation}
where $\mu$ is a parameter with mass dimension 1,
and $c$ is a dimensionless parameter of ${\cal O}(1)$.
Hereafter,
we use $M_{\rm pl}$ as the reduced Planck mass
($M_{\rm pl}\simeq 2.4\times 10^{18}{\rm\,GeV}$).
Higher order terms are expressed by the ellipsis.
Note that the parameter $\mu$ breaks the $R$ symmetry
since $Z$ has an $R$ charge of 0.
At the true minimum,
the VEV of $Z$ is assumed to be of the order of the Planck scale.
The parameter
$\mu$ is related to the gravitino mass as 
$|\mu|^2\simeq\left\langle|W_{\rm hid}|\right\rangle/M_{\rm pl}\simeq m_{3/2}M_{\rm pl}$.
The $F$-term of $Z$ is given by $|F_Z|\simeq m_{3/2}M_{\rm pl}$,
which implies spontaneous SUSY breaking.
The mass of the Polonyi field is of the order of $m_{3/2}$ for generic K\"ahler potentials.

The Polonyi model is attractive because of its simplicity
and has been studied extensively so far.
However,
such a singlet field causes cosmological difficulties~\cite{Coughlan:1983ci}.
In the early universe,
non-zero vacuum energy in the inflaton sector largely breaks SUSY,
which gives the Hubble induced mass term for the Polonyi field $Z$~\cite{Dine:1995kz,Dine:1995uk}.
Since $Z$ is singlet under any symmetry,
the minimum determined by the Hubble induced term
deviates from the true minimum.
The deviation is generically expected to be of the order of the fundamental energy scale of the supergravity,
which we assume to be the Planck scale.
Considering these SUSY breaking effects,
the Polonyi field evolves as follows.
During the inflation,
the Polonyi field $Z$ sits at the minimum determined by the Hubble induced term.
After the end of the inflation,
the Hubble parameter decreases to $m_Z$,
and then $Z$ begins to oscillate around the true minimum
with an amplitude of the order of the Planck scale.
The energy density of the oscillating field $Z$ scales as $a^{-3}$,
where $a$ is the scale factor,
and that of radiation scales as $a^{-4}$.
Thus,
the oscillating Polonyi field immediately dominates the universe after an inflaton decays into radiation.

When the Polonyi field starts its oscillation before the inflaton decays,
the Polonyi energy density-to-entropy ratio after the inflaton decay is given by
\begin{equation}
	\frac{\rho_Z}{s}
	\simeq\frac{T_{\rm inf}}{8}\left(\frac{z_0}{M_{\rm pl}}\right)^2,
	\label{eq:rho_Z/s}
\end{equation}
where $T_{\rm inf}$ denotes the temperature when the inflaton decays,
$s$ denotes the entropy density,
and $z_0$ represents the oscillation amplitude of $Z$,
which is expected to be of the order of the Planck scale.
With the decay rate of the Polonyi field,
\begin{equation}
	\Gamma_Z=\frac{d_Z}{8\pi}\frac{m_Z^3}{M_{\rm pl}^2},
	\label{eq:Gamma_Z}
\end{equation}
the decay temperature is given by
\begin{equation}
	T_{Z}
	=\left(\frac{90}{\pi^2g_*(T_{Z})}\right)^{1/4}\sqrt{\Gamma_ZM_{\rm pl}}
	\simeq 4{\rm\,MeV}\times d_Z^{1/2}
	\left(\frac{m_Z}{100{\rm\,TeV}}\right)^{3/2},
\end{equation}
where $d_Z$ is a numerical constant,
and $g_*(T_Z)$ is the effective number of degrees of freedom 
at a temperature of $T_Z$.
Here,
we used $g_*(T_{Z})=10.75$.
One can find that if $m_Z$ is smaller than ${\cal O}(100){\rm\,TeV}$,
the Polonyi field decays during and after the BBN.
In this case,
$\rho_Z/s\lesssim {\cal O}(10^{-14}\mathchar`-10^{-11}){\rm\,GeV}$ is required
in order not to upset the success of the BBN~\cite{Kawasaki:2004yh}.
From Eq.~(\ref{eq:rho_Z/s}),
one can find that
the constraint from the BBN cannot be avoided without dilution
even if $T_{\rm inf}\simeq{\cal O}(10){\rm\,MeV}$.
The required dilution factor $\Delta$ for $\rho_Z/s\lesssim 10^{-14}{\rm\,GeV}$ is as follows:
\begin{equation}
	\Delta\equiv\frac{s_fa^3}{s_ia^3}\gtrsim1.3\times10^{22}
	\left(\frac{T_{\rm inf}}{10^9{\rm\,GeV}}\right),
	\label{eq:Delta_1}
\end{equation}
where $s_f$ and $s_i$ denote the entropy density after and before the entropy production,
respectively.
Such huge entropy production,
however,
also dilutes pre-existing baryon asymmetry.
We then need an intricate model in which the baryon asymmetry is produced 
after the dilution~\cite{Stewart:1996ai,Jeong:2004hy,Kawasaki:2006py}.

There is another simple way to avoid the problem,
which we focus on in this paper.\footnote{For other solutions,
see Refs.~\cite{Linde:1996cx,Takahashi:2010uw,Takahashi:2011as,Nakayama:2011zy,Nakayama:2012mf,Harigaya:2013ns}.
}
When the Polonyi is as heavy as ${\cal O}(100){\rm\,TeV}$,
it decays before the onset of the BBN,
and the constraint becomes much milder~\cite{Kawasaki:2004yh}.
Even in this case,
however,
there is an incidental problem:
LSPs are abundantly produced from the decay of the Polonyi,
and the LSP density tends to exceed the observed DM density.
The abundance of the LSPs is given by
\begin{equation}
	Y_{\rm LSP}\equiv\frac{n_{\rm LSP}}{s}\simeq{\rm min}
	\left[\frac{\Gamma_Z}{\left\langle\sigma v\right\rangle s(T_Z)},
	\frac{N_{\rm LSP} n_Z(T_Z)}{s(T_Z)}\right],
\end{equation}
where $n_{\rm LSP}$ and $n_Z$ denote the number density of the LSP and the Polonyi field,
respectively.
$\left\langle\sigma v\right\rangle$ represents a thermally averaged cross section of the pair annihilation.
$N_{\rm LSP}$ is the averaged number of superparticles produced by the decay of one Polonyi field.
When the first term is relevant,
the pair annihilation between LSPs proceeds after the Polonyi decay,
and the relic LSP density is approximately proportional to $T_Z^{-1}$.
For example,
when the neutral wino is the LSP
with a mass of ${\cal O}(0.1\mathchar`-1){\rm\,TeV}$,
the decay temperature $T_{Z}$ is needed to be larger than
${\cal O}(1\mathchar`-10){\rm\,GeV}$~\cite{Moroi:2013sla} in order 
for the wino abundance not to exceed the observed DM density.\footnote{Reference \cite{Moroi:2013sla} has taken into account 
Sommerfeld effect and coannihilation among charged and neutral winos. 
}
This requires that $m_Z$ is larger than ${\cal O}(5000){\rm\,TeV}$.
Assuming that the gravitino mass is generically comparable to $m_Z$
and that the wino mass is generated by the anomaly-mediation,
such a heavy Polonyi mass is incompatible with the wino mass of ${\cal O}(0.1\mathchar`-1){\rm\,TeV}$.

In this paper,
we consider the case where $m_Z\simeq {\cal O}(100){\rm\,TeV}$ and the LSP density is diluted by entropy production.
As we will show,
it is possible to generate the baryon asymmetry before the dilution
since the required dilution factor is much smaller than Eq.~(\ref{eq:Delta_1}) for $m_Z=O(100)$ TeV.
The most probable candidate for baryogenesis is the Affleck-Dine mechanism
because it can create huge baryon number to survive the dilution.

\section{Affleck-Dine Mechanism without Superpotential}
\label{sec:Affleck-Dine}

The Affleck-Dine mechanism~\cite{Affleck:1984fy} is a promising candidate for the baryogenesis
in cosmological scenarios with dilution.
In this section,
we briefly explain the AD mechanism in the case
where the AD field does not appear in the superpotential.
We show that the resultant baryon number density is comparable to the number density of the Polonyi field.

The minimal SUSY Standard Model (MSSM)
contains a lot of flat directions which have no scalar potentials at the renormalizable level
and in SUSY limit~\cite{Gherghetta:1995dv}.
In the AD mechanism,
a flat direction with a $B-L$ charge
creates the baryon asymmetry.
We call it ``the Affleck-Dine field".
In the early universe,
SUSY breaking effects and non-renormalizable terms
affect its evolution.
In particular,
$A$-term scalar potentials violating $B-L$ global symmetry
rotate the AD field in the complex plane,
and can effectively generate $B-L$ number~\cite{Dine:1995kz}.
The $B-L$ asymmetry is converted into the baryon asymmetry 
through the sphaleron process~\cite{Kuzmin:1985mm,Khlebnikov:1988sr}.

In order to generate huge baryon number comparable to the number density of the Polonyi field,
the AD field value at the onset of its oscillation should be of the order of the Planck scale.
To obtain the large field value, we assume that the AD field does not appear in the superpotential.\footnote{For example,
a $U(1)_R$ symmetry can prohibit appearance of the AD field in the superpotential.
}
In this case,
SUSY breaking effects including the $A$-terms are provided by the K\"ahler potential.
We consider the following terms in the K\"ahler potential:\footnote{The second term in the second line is equivalent to a superpotential term suppressed by the gravitino mass.
}
\begin{eqnarray}
	{\cal L}_{\rm AD}&=&\int d^2\theta d^2\bar{\theta}\left[
	-3M_{\rm pl}^2\exp\left(-\frac{K}{3M_{\rm pl}^2}\right)\right] \nonumber \\
	&\supset&\int d^2\theta d^2\bar{\theta}\left[
	f_1|\Phi|^2
	+\left(f_2\frac{\Phi^n}{nM_{\rm pl}^{n-2}}+{\rm h.c.}\right)
	+f_3\frac{|\Phi|^4}{M_{\rm pl}^2}+\cdots\right],
	\label{eq:AD_superspace}
\end{eqnarray}
where $\Phi$ denotes the AD field,
and $f_i$ ($i=1,2,3$) is an arbitrary real function of $Z$ and $Z^{\dagger}$
which satisfies $f_i=f_i^{\dagger}$ for $i=1,3$.
From these terms,
the potential for the AD field $\Phi$ is given by
\begin{equation}
	V(\Phi)=m_{\phi}^2|\Phi|^2
	-\frac{m_{3/2}^2}{nM_{\rm pl}^{n-2}}\left(a_n\Phi^n+{\rm h.c.}\right)
	+c_4m_{3/2}^2\frac{|\Phi|^4}{M_{\rm pl}^2}+\cdots,
	\label{eq:AD_potential}
\end{equation}
where $m_{\phi}$ and $m_{3/2}$ denote the soft scalar mass of the AD field
and the gravitino mass,
respectively.
$a_n$ and $c_4$ are ${\cal O}(1)$ dimensionless parameters.

As we will discuss later,
$m_{\phi}$ is assumed to be smaller than $m_{3/2}$.
When $m_{\phi}\ll m_{3/2}$ and $c_4=0$,
there exists charge/color breaking minima smaller than the Planck scale
because of the relatively large $A$-terms~\cite{Kawasaki:2000ye}.
The AD field needs to avoid dropping the minima during its evolution,
which makes the AD mechanism less effective~\cite{Fujii:2001sn,Kawasaki:2006yb}.
The quartic term,
however,
lifts the potential near the Planck scale,
and the global minima disappear 
when its coefficient is positive.

In addition,
the AD field acquires the so-called Hubble induced terms
since the non-zero vacuum energy in the early universe largely violates SUSY~\cite{Dine:1995kz,Dine:1995uk}.
We assume that the AD field has a negative Hubble induced mass term
so that it takes a large field value at the onset of its oscillation:
\begin{equation}
	V_H=-c_HH^2|\Phi|^2,
\end{equation}
where $H$ is the Hubble parameter,
and $c_H$ is a positive dimensionless parameter of ${\cal O}(1)$.
Due to this negative mass term,
the AD field takes its field value of the order of the Planck scale until $H(t)\simeq m_{3/2}$.
When $H(t)\lesssim m_{3/2}$,
the position of the local minimum of the AD field,
which is determined by a balance between the negative Hubble induced mass term and the positive quartic term,
becomes smaller than the Planck scale.
Since the position is quickly driven towards the origin,
the AD field cannot track the local minimum and starts to roll down to the origin when $H(t)\simeq m_{3/2}$
(for details, see appendix~\ref{sec:appendix} and~\cite{Harigaya:2015hha}).

Let us estimate the produced baryon number density.
The evolution equation for the $B-L$ density is expressed as 
\begin{equation}
	\dot{n}_{B-L}+3Hn_{B-L}=2\beta{\rm Im}\left[\frac{\partial V}{\partial \Phi}\Phi\right],
\end{equation}
where $n_{B-L}$ expresses the $B-L$ density and $\beta$ denotes the $B-L$ charge 
of the AD field.
The dot denotes the time derivative.
The right-hand side is the source of the $B-L$ asymmetry.
By solving the evolution equation,
one can find that
the asymmetry is produced
most effectively at the onset of the oscillation ($H\simeq m_{3/2}$).
After that,
the AD field value decreases due to the expansion of the universe,
and $B-L$ violating effects become negligible.
The produced $B-L$ density is then estimated as
\begin{eqnarray}
	n_{B-L}(t_{\rm osc})&\sim&2\beta |a_n|\sin\left[n\theta_i+{\rm arg}(a_n)\right]
	\frac{m_{3/2}^2}{H_{\rm osc}}\frac{|\Phi_{\rm osc}|^n}{M_{\rm pl}^{n-2}} \nonumber \\
	&\equiv& \epsilon H_{\rm osc}|\Phi_{\rm osc}|^2,
\end{eqnarray}
where the subscripts of $_{\rm osc}$ show the values when the AD field starts to oscillate.
$\theta_i$ is the initial phase of the AD field.
$\epsilon$ is estimated as
\begin{equation}
	\epsilon\simeq 2\beta |a_n|\sin\left[n\theta_i+{\rm arg}(a_n)\right]
	\frac{m_{3/2}^2}{H^2_{\rm osc}}\frac{|\Phi_{\rm osc}|^{n-2}}{M_{\rm pl}^{n-2}}.
\end{equation}
As mentioned above,
the AD field starts to roll down to the origin from the field value of the order of the Planck scale
when $H(t)=H_{\rm osc}\simeq m_{3/2}$.
Therefore,
the $B-L$ asymmetry is estimated as 
$n_{B-L}(t_{\rm osc})\simeq \epsilon m_{3/2}|\Phi_{\rm osc}|^2$,
where $\epsilon\simeq{\cal O}(1)$
and $|\Phi_{\rm osc}|\simeq M_{\rm pl}$.

After the oscillation,
the AD field decays into radiation,
and the conserved $B-L$ asymmetry is converted into the baryon asymmetry
through the sphaleron effect~\cite{Kuzmin:1985mm,Khlebnikov:1988sr}.
The baryon asymmetry is related to the $B-L$ asymmetry as
\begin{equation}
	n_{B}=\frac{8}{23}n_{B-L},
\end{equation}
where $n_{B}$ expresses the baryon number density.
As mentioned in Sec.~\ref{sec:Polonyi},
entropy production is often necessary in order to avoid the overproduction of LSPs
by the decay of the Polonyi.
We consider the case where the density of the Polonyi field is diluted by the entropy production.
Then,
the baryon asymmetry is also diluted.
Assuming that the inflaton decays after the onset of the oscillation of the AD field,
the yield of the baryon number is estimated as
\begin{equation}
	Y_{B}\equiv\frac{n_B}{s}=
	\frac{8}{23}\frac{1}{\Delta}
	\left.\frac{3T_{\rm inf}n_{B-L}}{4\rho_{\rm inf}}\right|_{\rm osc}
	\simeq\frac{2}{23}\frac{\epsilon}{\Delta}\frac{T_{\rm inf}}{m_{3/2}}
	\left(\frac{|\Phi_{\rm osc}|}{M_{\rm pl}}\right)^2,
	\label{eq:Y_B}
\end{equation}
where $\rho_{\rm inf}$ denotes the energy density of the oscillating inflaton.
Note that the baryon number density is comparable to the density of the Polonyi field
because both the AD field and the Polonyi field simultaneously begin their oscillation
with the same amplitude of the order of the Planck scale.

Let us make a comment on Q-ball formation.
When the potential for the AD field is shallower than a quadratic potential,
the AD field fragments into non-topological solitons,
called Q-balls~\cite{Coleman:1985ki}, 
just after the onset of the oscillation~\cite{Kusenko:1997si,Enqvist:1998en,Kasuya:1999wu}.
Since Q-balls absorb the produced $B-L$ charge,
the formation of Q-balls
could significantly affect the estimation of the baryon asymmetry.
In our scenario,
the AD field value at the onset of the oscillation
is as large as the Planck scale.
Then,
the formed Q-balls may be too large to decay before the BBN
if Q-ball formation occurs,
which renders the AD mechanism ineffective.
Hence,
the beta function for the soft mass of the AD field may need to be positive
in order to prohibit the Q-ball formation.
This requires the AD field to involve scalar fields
which have large Yukawa couplings.

\section{Baryon-to-DM Ratio}
\label{sec:B-DM}

In this section,
we show that the baryon-to-DM ratio is simply given by the LSP mass
and the branching fraction of the decay of the the Polonyi into superparticles in our scenario.
We also explain that 
our scenario is realized in the so-called sequestering model~\cite{Inoue:1991rk,Randall:1998uk} with a (pseudo-)NGB, which can be identified with the QCD axion.

\subsection{Scenario}
\label{subsec:B-to-DM}

Before we calculate the baryon-to-DM ratio,
let us summarize our scenario.
When $H(t)\simeq m_Z\simeq m_{3/2}\simeq {\cal O}(100){\rm\,TeV}$,
both the Polonyi and the AD fields roll down to their origins
with the amplitudes of the order of the Planck scale.
At that time,
the AD field generates the $B-L$ asymmetry
which is later converted to the baryon asymmetry by the sphaleron process.
Then,
the entropy production occurs
and dilutes both the Polonyi density and the baryon asymmetry.
After the dilution,
the Polonyi field decays into superparticles
which consequently decay into LSPs
before the epoch of the BBN.
Thus,
the DM density is determined by the abundance of the nonthermally produced LSPs,
assuming that thermal relic density of LSPs is negligible.

First,
we estimate the DM abundance.
From Eq.~(\ref{eq:rho_Z/s}),
the LSP-to-entropy ratio is estimated as
\begin{equation}
	\frac{\rho_{\rm LSP}}{s}=m_{\rm LSP}\frac{2{\rm Br}_{\rm SUSY}}{\Delta}\frac{n_{Z}}{s_i}
	\simeq \frac{{\rm Br}_{\rm SUSY}}{\Delta}\frac{T_{\rm inf}m_{\rm LSP}}{4m_Z}\left(\frac{z_0}{M_{\rm pl}}\right)^2,
	\label{eq:Y_LSP}
\end{equation}
where ${\rm Br}_{\rm SUSY}$ denotes a branching fraction for the Polonyi decay
into two superparticles,\footnote{Gravitinos are not produced from the Polonyi decay 
assuming that the decay is kinematically forbidden
($m_Z<2m_{3/2}$).
The abundance of gravitinos produced during the reheating 
becomes negligible after the dilution.
}
and $m_{\rm LSP}$ denotes the LSP mass.
The number of the produced superparticles
is almost equal to that of the LSPs due to the $R$-parity conservation.
Note that the pair annihilation between LSPs is not efficient
since the Polonyi field decays after the dilution.
We assume that decay products other than the LSPs
do not contribute to the DM abundance.
Hence,
the DM abundance is obtained by the nonthermally produced LSPs.

Next,
let us compare the DM abundance with the baryon asymmetry 
created by the AD mechanism.
The ratio of the density of the Polonyi field to the $B-L$ number
remains the same after they begin their oscillations
since the densities of both components decrease as $a^{-3}$.
From Eqs.~(\ref{eq:Y_B}) and (\ref{eq:Y_LSP}),
we obtain the following relation:
\begin{eqnarray}
	\frac{\Omega_B}{\Omega_{\rm LSP}}&=&\frac{8}{23}\frac{\epsilon}{\rm Br_{\rm SUSY}}
	\frac{m_pm_Z}{m_{\rm LSP}m_{3/2}}\left(\frac{|\Phi_{\rm osc}|}{z_0}\right)^2 \nonumber \\
	&\simeq&0.33\epsilon\left(\frac{10^{-3}}{\rm Br_{\rm SUSY}}\right)
	\left(\frac{1{\rm\,TeV}}{m_{\rm LSP}}\right)\left(\frac{|\Phi_{\rm osc}|}{z_0}\right)^2,
	\label{eq:Omega_B/Omega_LSP}
\end{eqnarray}
where
$m_p$ represents the proton mass
($m_p\simeq 0.938{\rm\,GeV}$).
Here,
we assume $m_Z\simeq m_{3/2}$.
Note that $z_0$ and $|\Phi_{\rm osc}|$ are of the order of the Planck scale
and that $\epsilon$ is of ${\cal O}(1)$.
One can find that 
the baryon-to-DM ratio is determined by the LSP mass and the branching fraction
of the decay of the Polonyi into superparticles.
Assuming that the LSP mass is of ${\cal O}(1){\rm\,TeV}$,
${\rm Br}_{\rm SUSY}$ is
required to be of ${\cal O}(10^{-3})$
in order to realize the observed value, $\Omega_{B}/\Omega_{\rm DM}\simeq 0.18$.

Our scenario needs the entropy production
(see appendix~\ref{sec:entropy} for the realization by thermal inflation). 
Let us estimate the required amount of the entropy production.
The ratio of the observed DM density to the entropy density is given by~\cite{Agashe:2014kda}
\begin{equation}
	\frac{\rho^{\rm (obs)}_{\rm DM}}{s_0}\simeq 4.4\times10^{-10}{\rm\,GeV},
	\label{eq:DM/s}
\end{equation}
where $\rho^{\rm (obs)}_{\rm DM}$ denotes the observed DM energy density,
and $s_0$ denotes the present entropy density.
Comparing Eq.~(\ref{eq:Y_LSP}) with Eq.~(\ref{eq:DM/s}),
the dilution factor of the required entropy production is estimated as
\begin{equation}
	\Delta\simeq 1.9\times10^{12}
	\left(\frac{T_{\rm inf}}{10^9{\rm\,GeV}}\right)
	\left(\frac{m_{\rm LSP}}{1{\rm\,TeV}}\right)\left(\frac{\rm Br_{\rm SUSY}}{10^{-3}}\right)
	\left(\frac{300{\rm\,TeV}}{m_Z}\right)
	\left(\frac{z_0}{M_{\rm pl}}\right)^2,
	\label{eq:Delta}
\end{equation}
where we assume that the inflaton decays after the Polonyi and the AD fields start to oscillate.
This is the case for $T_{\rm inf}\lesssim10^{12}{\rm\,GeV}(m_Z/300{\rm\,TeV})^{1/2}$.
$m_Z$ should be of ${\cal O}(100){\rm\,TeV}$
in order to relax the constraint from the BBN.
Comparing Eq.~(\ref{eq:Delta}) with Eq.~(\ref{eq:Delta_1}),
one can find that the required dilution factor is much smaller than the 
case of $m_Z\lesssim {\cal O}(100){\rm\,TeV}$.
Note that the dilution factor is estimated assuming that
the entropy production occurs before the Polonyi decays.

\subsection{Sequestering Model and Decay Process of the Polonyi Field}
\label{subsec:sequestering}

When the Polonyi field decays through dimension 5 operators 
suppressed by the Planck scale (see Eq.~(\ref{eq:Gamma_Z})),
the branching fraction of the decay into superparticles is generally
comparable to that into the SM particles~\cite{Endo:2006ix}.
In order to suppress the branching fraction 
into superparticles (${\rm Br}_{\rm SUSY}\simeq{\cal O}(10^{-3})$),
we consider the so-called sequestering model~\cite{Inoue:1991rk,Randall:1998uk},\footnote{The sequestering model has been introduced
in the context of extra dimension~\cite{Randall:1998uk}.
It is also realized in a four dimensional strongly coupled CFT~\cite{Luty:2001jh,Ibe:2005pj}.
}
in which the SUSY breaking sector is sequestered from the visible sector.

The K\"ahler potential and the superpotential are given by
\begin{equation}
	K=-3M_{\rm pl}^2\log\left[1-\frac{f_{\rm vis}+f_{\rm hid}}{3M_{\rm pl}^2}\right],~~~
	W=W_{\rm vis}+W_{\rm hid},
	\label{eq:Kahler_super}
\end{equation}
where the subscripts of $_{\rm vis}$ and $_{\rm hid}$ denote the visible 
and the hidden sectors (SUSY breaking sector),
respectively.
We also assume that
the SM gauge sector does not directly couple to the hidden sector:
\begin{equation}
	{\cal L}_{\rm gauge}=\int d^2\theta\left[\tau_{\rm vis}{\cal W}^a{\cal W}_{a}+{\rm h.c.}\right],
\end{equation}
where ${\cal W}^a$ denotes field strength supermultiplets of the visible SM gauge sector,
and $\tau_{\rm vis}$ is a holomorphic function which depends only on visible sector fields.

In this setup,
gaugino masses vanish at the tree level
because the Polonyi field does not appear in the gauge sector.
The quantum corrections to the gaugino masses
arise only at loop-suppressed level,
which mainly come from the anomaly mediation~\cite{Randall:1998uk,Giudice:1998xp}.
Then,
the lightest gaugino is the neutral wino
with a mass of ${\cal O}(1){\rm\,TeV}$
when $m_{3/2}\simeq{\cal O}(100){\rm\,TeV}$.
This is compatible with our scenario
with the neutral wino LSP.

Soft scalar masses also vanish at the tree level
when the K\"ahler potential is given by Eq.~(\ref{eq:Kahler_super}).
They acquire
loop-suppressed contribution
from the anomaly mediation~\cite{Randall:1998uk,Giudice:1998xp},
Planck-suppressed interactions~\cite{Antoniadis:1997ic} and so on.
If the MSSM scalars acquire their masses only from the anomaly mediation,
slepton masses would become negative.
This is problematic in terms of the phenomenology.
Thus,
there should be other sources to give them
positive masses.
One of such candidates is
one loop correction from the Planck-suppressed interactions~\cite{Randall:1998uk,Antoniadis:1997ic}.
When a cut-off scale is taken around the gravitational scale,
one loop correction can exceed the anomaly-mediated masses
which appear at two loop level.\footnote{When the one loop correction determines scalar masses,
the mass spectra of MSSM scalar particles become UV sensitive,
which is contrary to the anomaly-mediated masses.
Thus,
we lose a solution to the SUSY FCNC problem
unless the universality condition is imposed at the UV scale.
There also exists other UV insensitive models 
which solve the negative slepton mass problem~\cite{Nelson:2002sa,Hsieh:2006ig,Gupta:2012gu,Pomarol:1999ie,Abe:2001cg}.
}
In this case,
sfermion masses are of ${\cal O}(10){\rm\,TeV}$
when $m_{3/2}\simeq{\cal O}(100){\rm\,TeV}$.\footnote{The lightest Higgs boson mass acquires
radiative corrections from stop one loop diagrams~\cite{Okada:1990vk}.
Stop mass of ${\cal O}(10){\rm\,TeV}$ is compatible with the relatively heavy observed Higgs boson mass of 126{\rm\,GeV}.
}

When the soft masses and the supersymmetric masses ($\mu$ term) 
of the Higgs fields are 
of ${\cal O}(10){\rm\,TeV}$ and of ${\cal O}(1){\rm\,TeV}$ respectively,
the $B\mu$ term ($\sim\mu m_{3/2}$) is comparable to the scalar masses,
which leads to the successful electroweak gauge symmetry breaking.
The higgsino with mass of $\mu\sim{\cal O}(1){\rm\,TeV}$ could be the LSP
instead of the neutral wino.
When the soft masses are of ${\cal O}(1){\rm\,TeV}$,
the $B\mu$ term is generally too large to realize the electroweak symmetry breaking.
In the Next-to-MSSM~\cite{Ellwanger:2009dp},
however,
the supersymmetric Higgs mass term is generated as the breaking term of the scale invariance,
and the (effective) $B\mu$ term appears at the loop-suppressed level.

Since the SUSY breaking sector is now sequestered from the AD field,
the functions $f_i$ ($i=1,2,3$) in Eq.~(\ref{eq:AD_superspace}) do not contain the Polonyi field.\footnote{In the early universe,
the inflaton sector breaks the SUSY,
which generates the Hubble induced mass term.
In order to generate the negative Hubble induced mass term for the AD field,
the inflaton sector should not be sequestered from the visible sector.
}
Even in this case,
the potential for the AD field involves
the holomorphic $A$-terms and the quartic term of ${\cal O}(m_{3/2}^2)$
due to the explicit breaking of the conformal symmetry.
By requiring that the vacuum energy vanishes,
the coefficients in Eq.~(\ref{eq:AD_potential}),
$a_n$ and $c_4$,
are estimated as $a_n\simeq -f_2(n-1)$ and $c_4\simeq f_3$
when $f_1=1$.
Note that the estimated values contain uncertainties of ${\cal O}(1)$.

Let us consider the decay process of the Polonyi field (for details, see~\cite{Endo:2006ix,Nakamura:2007wr}).
Firstly,
the Polonyi field generally decays into 2 gravitinos at the tree level when $m_Z>2m_{3/2}$.
This decay process is incompatible with our scenario
since the branching fraction of the decay of the Polonyi into superparticles is required to be of ${\cal O}(10^{-3})$.
Hence,
we assume that 
the decay into 2 gravitinos is kinetically forbidden ($m_Z<2m_{3/2}$).

The decay into matter scalars comes from the kinetic terms for the sequestered potential:
${\cal L}_{K}=g_{ij^*}\partial_{\mu}\phi^i\partial^{\mu}\phi^{*j}$,
where $\phi^i$ denotes the matter scalar fields,
and $g_{ij^*}=\frac{\partial^2K}{\partial \phi^i\partial\phi^{*j}}$.
The kinetic terms are converted into the following form up to a total derivative:
${\cal L}\sim\frac{Z}{M_{\rm pl}}\phi^i\partial^2\phi^{*j}$.
Using the equation of motion,
interaction terms from the kinetic terms are proportional to the scalar mass squared.
Thus,
the branching fraction of the decay mode $Z\to\phi^i\phi^{*j}$ is suppressed
by a factor of ${\cal O}(m_{\phi}^4/m_{Z}^4)\sim {\cal O}(10^{-4}\mathchar`-10^{-5})$
when the scalar mass is smaller than the Polonyi mass.
The Polonyi field also decays into matter scalar fields through one-loop diagrams by Planck-suppressed interactions,
but the rates of these decays are the same order with that of the tree-level decay.
Similarly,
the branching fraction into matter fermions is proportional to fermion mass squared
and is negligible.
The decay rate into higgsinos with masses of $\mu\sim{\cal O}(1){\rm\,TeV}$
is the same order with that into matter scalar fields
since it is suppressed by a factor of ${\cal O}(\mu^2/m_{Z}^2)\sim{\cal O}(10^{-4})$.

Decay into three-body final states is suppressed for the sequestered potential.
In general,
the decay of $Z\to\phi^i\chi^j\chi^k$,
where $\chi^i$ denotes the matter fermions,
occurs through the following interaction:
\begin{equation}
	{\cal L}_{\rm three}=-\frac{1}{2}e^{\frac{K}{2M_{\rm pl}^2}}
	\left(\frac{K_Z}{M_{\rm pl}^2}W_{ijk}-3\Gamma^{l}_{Zi}W_{jkl}\right)
	Z\phi^i\chi^j\chi^k+{\rm h.c.},
\end{equation}
where the subscripts represent the derivative by the scalar fields,
and $\Gamma^i_{jk}=g^{il^*}g_{jl^*k}$.
One can find that this term vanishes
if the K\"ahler potential is given by the form of Eq.~(\ref{eq:Kahler_super}).
For the same reason,
the decay of $Z\to\phi^i\phi^j\phi^k$ does not occur,
either.

Since the Polonyi field is not directly coupled with the gauge sector,
it does not decay into gauge bosons and gauginos at the tree level.
However,
it can decay into them
through the anomaly-mediated effects.
When the mass of the Polonyi field is dominated by a supersymmetric mass term,
the interaction terms between $Z$ and the gaugino $\lambda$ are given by~\cite{Endo:2007ih}
\begin{equation}
	{\cal L}_{\rm anomaly}=\frac{\alpha_g b_0m_Z}{24\pi M_{\rm pl}}
	\frac{K_Z}{M_{\rm pl}}Z^*\lambda\lambda+{\rm h.c.},
	\label{eq:L_anomaly}
\end{equation}
where $\alpha_g=g^2/4\pi$ represents a gauge coupling constant,
and $b_0=3T_G-T_R$ is the coefficient of the beta function.
Since the SUSY breaking mass term is comparable to the supersymmetric mass term,
the interaction terms are deviated from Eq.~(\ref{eq:L_anomaly}) by ${\cal O}(1)$.
From Eq.~(\ref{eq:L_anomaly}),
the decay rate is estimated as~\cite{Endo:2007ih}
\begin{equation}
	\Gamma(Z\to2\lambda)\simeq\frac{N_g\alpha_g^2b_0^2}{4608\pi^3}
	\frac{|K_Z|^2}{M_{\rm pl}^2}\frac{m_Z^3}{M_{\rm pl}^2},
	\label{eq:anomaly_decay}
\end{equation}
where $N_g$ is the number of gauginos.
The decay rate of $Z$ into 2 gauge bosons is also the same as Eq.~(\ref{eq:anomaly_decay}).
The most important process is the decay into gluons and gluinos.
We can estimate its rate by using $N_g=8$ and $b_0=3$.

In summary,
the Polonyi field mainly decays into gluinos and gluons through the anomaly-mediated effects
for the sequestered K\"ahler potential.
If it is the leading process,
however,
the Polonyi field becomes long-lived,
and the constraint from the BBN is again severe even with $m_Z\simeq{\cal O}(100){\rm\,TeV}$.
We need some other efficient decay processes.
Note that those decay processes should not yield large DM abundance.
As a suitable decay process, we consider the decay of the Polonyi field into a (pseudo-)NGB.
To be specific,
we introduce the QCD axion~\cite{Peccei:1977hh,Peccei:1977ur,Weinberg:1977ma,Wilczek:1977pj}.
The axion is a pseudo-NGB 
associated with the spontaneous breaking of the Peccei-Quinn (PQ) symmetry
and appears as a phase direction of the PQ field:
\begin{equation}
	P= v_{\rm PQ}\exp\left(\frac{s+ia}{\sqrt{2}v_{\rm PQ}}\right),
	\label{eq:PQ_VEV}
\end{equation}
where $P$, $s$ and $a$ denote the PQ field, the saxion field and the axion field,
respectively.
$v_{\rm PQ}$ is the VEV of the PQ field.

Let us assume that the PQ field also belongs to the visible sector,
which is natural as the PQ field must directly couple to SM charged particles.
The PQ field interacts with the Polonyi field 
through the kinetic terms as follows:
\begin{equation}
	{\cal L}\sim\frac{Z}{M_{\rm pl}}P\partial^2 P^{*}+{\rm h.c.}.
	 \label{eq:Pdel^2P}
\end{equation}
After the PQ symmetry breaking,
the axion field appears as a massless direction as Eq.~(\ref{eq:PQ_VEV}).
Expanding of Eq.~(\ref{eq:Pdel^2P}) in terms of the saxion and axion
leads to mixing of the kinetic terms between the Polonyi field and (s)axion.
In order to estimate the rate of the Polonyi decay into axions,
we need to diagonalize the kinetic terms and
transform the bases into mass eigenstates.
We then obtain the following interactions:
\begin{equation}
	{\cal L}=c_R\frac{z_R}{\sqrt{2}M_{\rm pl}}\partial_{\mu}\hat{a}\partial^{\mu}\hat{a}
	+c_I\frac{z_I}{\sqrt{2}M_{\rm pl}}\partial_{\mu}\hat{a}\partial^{\mu}\hat{a},
	\label{eq:no_mass_suppress}
\end{equation}
where $c_R$ and $c_I$ are coefficients of ${\cal O}(1)$,
and $z_R$ and $z_I$ denote a real component 
and an imaginary component of the Polonyi field ($Z=\frac{1}{\sqrt{2}}(z_R+iz_I$)),
respectively. 
Here,
we used $\hat{a}$ to show the mass eigenstate of the massless direction.

The rate of the Polonyi decay into axions\footnote{The decay products of the Polonyi field could contain the axino
which is a superpartner of the axion.
Since the decay into axinos could lead to the overproduction of LSPs,
we assume that such a decay process is kinematically forbidden,
$2m_{\tilde{a}}>m_Z$,
where $m_{\tilde{a}}$ represents the axino mass.
}
is estimated as
\begin{equation}
	\Gamma(Z\to 2~{\rm axions})=\frac{c_a^2}{64\pi}\frac{m_Z^3}{M_{\rm pl}^2},
	\label{eq:tree-level}
\end{equation}
where we define $c_a$ as $c_a^2\equiv c_R^2+c_I^2$.
The Polonyi decay temperature $T_Z$ is estimated as
\begin{equation}
	T_{Z}\simeq7.1{\rm\,MeV}c_a\left(\frac{m_Z}{300{\rm\,TeV}}\right)^{3/2}.
\end{equation}
Note that the Polonyi density does not dominate the universe at its decay
since we assume that the dilution occurs 
before the decay.
Even when the Polonyi field is a subdominant component of the universe,
it must decay before the BBN in order not to destroy synthesized light elements.
Hence,
the Polonyi should be as heavy as ${\cal O}(100){\rm\,TeV}$.

The rate of the loop-suppressed decay (Eq.~(\ref{eq:anomaly_decay}))
is much smaller than that of the tree-level decay (Eq.~(\ref{eq:tree-level})),
and we obtain the branching fraction of the Polonyi decay into superparticles as
\begin{equation}
	{\rm Br}_{\rm SUSY}=\frac{\Gamma(Z\to 2~{\rm superparticles})}{\Gamma(Z\to 2~{\rm axions})}\sim1\times10^{-3}.
	\label{eq:Br_SUSY}
\end{equation}
Since the axion mass is typically much smaller than the LSP mass,
the abundance of axions produced from the Polonyi decay is negligible compared with the LSP abundance.\footnote{We also assume that the density of the coherent oscillation of the axion field 
does not exceed the observed DM density,
which implies $v_{\rm PQ}/N_{\rm DW}\sim 10^{9\mathchar`-13}{\rm\,GeV}$.
}
Therefore,
the DM abundance is determined by the abundance of the decay products of
the suppressed decay into superparticles.
The axion also gives a negligible contribution to the dark radiation.
From Eqs.~(\ref{eq:Omega_B/Omega_LSP}) and (\ref{eq:Br_SUSY}),
it is found that the observed baryon-to-DM ratio of $\Omega_B/\Omega_{\rm DM}\simeq0.18$
is explained in the sequestering model with the (pseudo-)NGB.

\section{Summary and Discussions}
\label{sec:summary}

In this paper,
we have considered the cosmological consistent scenario
in the presence of the heavy Polonyi field.
When the Polonyi field is heavier than ${\cal O}(100){\rm\,TeV}$,
it decays before the BBN,
and the constraint from the BBN becomes much milder.
LSPs produced by the decay of the Polonyi are diluted by entropy production.
With the entropy production,
the most promising candidate for the baryogenesis is the AD mechanism.

In our scenario,
the DM is explained by the LSPs produced from the decay of the Polonyi field,
and the baryon asymmetry is created by the AD mechanism without superpotential.
We have shown that the baryon-to-DM ratio is simply determined by
the LSP mass ($m_{\rm LSP}$) and the branching fraction of
the decay of the Polonyi into DM (${\rm Br_{\rm SUSY}}$).
The observed ratio $\Omega_B/\Omega_{\rm DM}\simeq 0.18$
is realized for the LSP mass of ${\cal O}(1){\rm\,TeV}$
and the branching fraction of ${\cal O}(10^{-3})$.
The Polonyi density is diluted before its decay by the entropy production,
for example,
by thermal inflation (see appendix~\ref{sec:entropy}).

In general,
the branching fraction,
${\rm Br}_{\rm SUSY}$,
is of ${\cal O}(1)$ when the Polonyi field is connected with the observable sector
through non-renormalizable interactions.
In order to suppress ${\rm Br}_{\rm SUSY}$ $(\simeq{\cal O}(10^{-3}))$,
we have considered the sequestering model with a (pseudo-) NGB,
which can be identified with the QCD axion.
In this model,
the Polonyi field decays into NGBs at the tree level,
which do not contribute to the DM abundance.
On the other hand,
it decays
into superparticles mainly through anomaly-induced interactions
and hence is suppressed compared with the decay into NGBs.

Let us comment on the implication of our study to
the cosmological moduli problem~\cite{Banks:1993en,deCarlos:1993jw}.
There may exist other singlet scalar fields called ``moduli fields",
which are motivated from UV physics such as the superstring theory.
The late-time decay of those fields could also cause cosmological difficulties:
As is the case with the Polonyi field,
the abundance of nonthermally produced LSPs could exceed the observed DM density
even when the moduli field is heavy enough to decay before the BBN.
In that case,
the entropy production is needed.
However,
enough baryon asymmetry cannot be produced even with the AD mechanism
because the branching fraction ${\rm Br}_{\rm SUSY}$ is generally of the order unity
for generic moduli fields (see Eq.~(\ref{eq:Omega_B/Omega_LSP})).
Note that the AD mechanism discussed in this paper works the most effectively.
The cosmological moduli problem is very severe from the viewpoint of baryogenesis.

\section*{Acknowledgments}
This work is supported by Grant-in-Aid for Scientific research 
from the Ministry of Education, Science, Sports, and Culture
(MEXT), Japan,
No.~15H05889 (M.K.) and No.~25400248 (M.K.), 
JSPS Research Fellowships for Young Scientists (No.~25.8715 (M.Y.)), 
World Premier International Research Center Initiative (WPI Initiative), MEXT, Japan (M.K. and M.Y.),
the Program for the Leading Graduate Schools, MEXT, Japan (T.H. and M.Y.),
the Director, Office of Science, Office of High Energy and Nuclear Physics, of the U.S. Department of Energy under Contract DE-AC02-05CH11231 (K.H.), by the National Science Foundation under grants PHY-1316783 and PHY-1521446 (K.H.).

\appendix
\renewcommand{\theequation}{\Alph{section}.\arabic{equation}}
\setcounter{equation}{0}

\section{Evolution of the AD Field}
\label{sec:appendix}

Following Ref.~\cite{Dine:1995kz,Harigaya:2015hha},
we make a remark about the evolution of the AD field in the potential given by Eq.~(\ref{eq:AD_potential}).
In particular,
we pay attention to the onset of the oscillation of the AD field.

Including the negative Hubble induced mass term,
the potential for the radial component of the AD field,
$\phi\equiv|\Phi|/\sqrt{2}$,
is given by
\begin{equation}
	V(\phi)\simeq -\frac{c_H}{2}H^2\phi^2+c_4m_{3/2}^2\frac{\phi^4}{4M_{\rm pl}^2}+\cdots,
	\label{eq:V_phi}
\end{equation}
where we omit the soft SUSY breaking mass term
assuming that $H\gg m_{\phi}$.
When $H\gtrsim m_{3/2}$,
the AD field sits at some value of the order of the Planck scale. 
This is because higher dimensional operators lift the scalar potential.
When $m_{\phi}\lesssim H \lesssim m_{3/2}$,
local minimum determined by the potential is given by
\begin{equation}
	\phi_{\rm min}\simeq\sqrt{\frac{c_H}{c_4}}\frac{M_{\rm pl}H}{m_{3/2}}.
\end{equation}
One can find that $\phi_{\rm min}$ decreases as $a^{-3/2}$ from the Planck scale
during the epoch of the inflaton oscillation.
When $\phi_{\rm min}$ starts to decrease when $H\simeq m_{3/2}$,
the AD field cannot track the local minimum
and then starts to roll down to the origin.

In order to look more closely at the onset of the oscillation of the AD field,
we write down the equation of motion for $\phi$ as
\begin{equation}
	\ddot{\phi}+3H\dot{\phi}-c_HH^2\phi+c_4m^2_{3/2}\frac{\phi^3}{M_{\rm pl}^2}=0.
\end{equation}
Using the number of e-folding $N\equiv\ln(a/a_i)$ as a time variable,
this is rewritten as
\begin{equation}
	\frac{d^2}{dN^2}\phi+\frac{3}{2}\frac{d}{dN}\phi-c_H\phi+\frac{c_4m_{3/2}^2}{M_{\rm pl}^2H^2}\phi^3=0,
\end{equation}
where we take $a_i$ as the scale factor when $H_i\simeq m_{3/2}$.
Rescaling the AD field value $\phi$ as
\begin{equation}
	\psi\equiv\frac{\phi}{\phi_{{\rm min},i}}e^{\frac{3N}{2}},~~~
	\phi_{{\rm min},i}=\sqrt{\frac{c_H}{c_4}}\frac{M_{\rm pl}H_i}{m_{3/2}},
\end{equation}
we can eliminate the dependence on the time variable in the coefficients.
In terms of $\psi$,
the equation of motion is rewritten as
\begin{equation}
	\frac{d^2}{dN^2}\psi-\frac{3}{2}\frac{d}{dN}\psi-c_H\psi+c_H\psi^3=0.
\end{equation}
Note that the coefficient of the friction term is negative.
This implies that the AD field cannot track the local minimum
and starts to oscillate around the origin.\footnote{Reference \cite{Harigaya:2015hha} numerically confirms this behavior in the context of the evolution of the PQ field.
}
$B-L$ asymmetry is effectively produced at the onset of the oscillation
($H\simeq m_{3/2}$).
After that,
the asymmetry is conserved since
the amplitude of the AD field decreases,
and the $U(1)_{B-L}$ symmetry breaking terms become ineffective.

\setcounter{equation}{0}

\section{Thermal inflation}
\label{sec:entropy}

Our scenario needs the entropy production by some mechanism
in order to dilute the Polonyi density
and the baryon asymmetry,
which requires another component
to dominate the energy density of the universe.
Its energy density must be constant or decrease more slowly than that of the oscillating Polonyi field.
In this appendix,
we introduce thermal inflation~\cite{Lyth:1995ka}
as one example of dilution mechanisms.

Let us introduce a specific model of the thermal inflation.
The thermal inflation is a short epoch of accelerated expansion of the universe 
at a low-energy scale.
This mechanism requires a scalar field corresponding to a flat direction
at the renormalizable level,
which is called ``the flaton".
We assume an approximate $Z_4$ symmetry
and the superpotential given by
\begin{equation}
	W=\frac{\lambda_X}{4M_{\rm pl}}X^4
	+g_{\xi}X\xi \bar{\xi},
	\label{eq:W_X}
\end{equation}
where $\lambda_X$ and $g_{\xi}$ are dimensionless coupling constants,
and $X$ is the supermultiplet of the flaton field with a $Z_4$ charge of 1.
$\xi$ and $\bar{\xi}$ are massless $SU(3)_C$ gauge charged fields.
Note that $X$ is singlet under the SM gauge symmetry.
The massless gauge charged fields interact with thermal bath,
which generates the thermal mass term for $X$.
Here,
we ignore higher dimensional terms 
since we focus on the field value much smaller than the Planck scale.

Including SUSY breaking effects and the thermal mass term,
the potential is given by
\begin{equation}
	V(X)=V_0+(c_TT^2-m_0^2)|X|^2
	+\frac{m_{3/2}}{4M_{\rm pl}}\left(\lambda_XX^4+{\rm h.c.}\right)
	+\frac{|\lambda_X|^2}{M_{\rm pl}^2}|X|^6+\cdots,
	\label{eq:potential_X}
\end{equation}
where $V_0$ represents the vacuum energy which causes the thermal inflation,
and $c_T$ is a coefficient of the order of the square of $g_{\xi}$.
Hereafter,
we assume that $X$ has a tachyonic mass term around the origin ($m_0>0$).
We also assume that the flaton sector is sequestered from the SUSY breaking sector
and that $m_0$ is smaller than $m_{3/2}\simeq{\cal O}(100){\rm\,TeV}$.

The evolution of the flaton $X$ is as follows:
We assume that $X$ obtains a positive Hubble induced mass term  
during the primordial inflation.
Then,
$X$ is expected to sit around the origin just after the inflation.
Even when $H \lesssim m_0$,
$X$ can be trapped around the origin due to the thermal mass term,
and then the vacuum energy $V_0$ causes the accelerated expansion of the universe
at a low-energy scale.
The thermal inflation lasts until 
the temperature decreases to the critical value of $T_c\simeq c_T^{-1/2}m_0$.
After the end of the thermal inflation,
$X$ starts to roll down to the true minimum due to the negative mass term.
When $H$ decreases to the decay rate of $X$,
$X$ decays into radiation with huge entropy production,
and the radiation dominated universe is realized.

The true minimum of 
the potential is determined by the $A$-term
and non-renormalizable terms
when $m_0\ll m_{3/2}$.
The flaton VEV at present
is given by
\begin{eqnarray}
	\left\langle |X|\right\rangle\equiv M\simeq
	\left(\frac{m_{3/2}M_{\rm pl}}{3|\lambda_X|}\right)^{1/2} 
	\simeq 4.9\times 10^{11}{\rm\,GeV}|\lambda_X|^{-1/2}\left(\frac{m_{3/2}}{300{\rm\,TeV}}\right)^{1/2}.
\end{eqnarray}
Hereafter,
we assume that $\lambda_X$ is of ${\cal O}(1)$.
One can find that the VEV 
is much larger than the electroweak scale.
Therefore,
the $SU(3)_C$ charged matter $\xi$ ($\bar{\xi}$)
with mass of the order of $g_{\xi}\left\langle |X|\right\rangle$
is expected to be much heavier than the electroweak scale at present.\footnote{Note that the thermal mass term exists only when $g_{\xi}\left\langle |X|\right\rangle\ll T$.
When $X$ is trapped at the origin
during the thermal inflation,
the charged matters $\xi$ ($\bar{\xi}$) behave as relativistic particles in thermal bath.
}
$V_0$ is determined as follows by requiring that
the vacuum energy vanishes at the true minimum:
\begin{equation}
	V_0\simeq\frac{1}{18}M^2m_{3/2}^2
	\simeq1.2\times 10^{33}{\rm\,GeV}^4|\lambda_X|^{-1}\left(\frac{m_{3/2}}{300{\rm\,TeV}}\right)^3.
\end{equation}
When $X$ has its large VEV,
it is decomposed as
\begin{equation}
	X=\left[M+\frac{\chi}{\sqrt{2}}\right]
	\exp\left(i\frac{a_{\chi}}{\sqrt{2}M}\right),
\end{equation}
where $\chi$ and $a_{\chi}$ are canonically normalized real scalar fields.
We obtain a mass of the radial component $\chi$ around the true minimum:
\begin{equation}
	m_{\chi}\simeq\sqrt{\frac{2}{3}}m_{3/2}\simeq 240{\rm\,TeV}
	\left(\frac{m_{3/2}}{300{\rm\,TeV}}\right).
\end{equation}
If the $R$ symmetry was not broken,
the superpotential for $X$ would have $R$ symmetry
and the phase component $a_{\chi}$ would be a massless $R$ axion.
However,
the $R$ symmetry breaking, namely the non-zero VEV of the superpotential, generates the $R$ symmetry breaking $A$-terms
and the phase component $a_{\chi}$ also obtains its mass as $m_{a_{\chi}}\simeq\sqrt{4/3}m_{3/2}$.\footnote{Since $m_{\chi}<2m_{a_{\chi}}$,
the decay of $\chi$ into the phase components is forbidden.
}

After the end of the thermal inflation,
the energy density of the oscillating flaton field dominates that of the universe.
Thus,
the reheating occurs 
when $H$ decreases to the decay rate of the flaton $\chi$.
It mainly decays into gluons through one loop diagrams
of $\xi$ and $\bar{\xi}$.
The decay rate is given by~\cite{Asaka:1997rv,Asaka:1999xd}
\begin{equation}
	\Gamma(\chi\to 2g)=\frac{1}{4\pi}\left(\frac{\alpha_s}{4\pi}\right)^2
	\frac{m_{\chi}^3}{M^2}.
\end{equation}
The reheating temperature $T_{\rm RH}^{\chi}$ is estimated as
\begin{eqnarray}
	T_{\rm RH}^{\chi}&\simeq&\left(\frac{90}{\pi^2g_*(T_{\rm RH}^{\chi})}\right)^{1/4}
	\sqrt{\Gamma(\chi\to2g)M_{\rm pl}} \nonumber \\
	&\simeq&470{\rm\,GeV}|\lambda_X|^{1/2}\left(\frac{m_{3/2}}{300{\rm\,TeV}}\right)
	\left(\frac{\alpha_s}{0.1}\right),
\end{eqnarray}
where we use $g_*(T_{\rm RH}^{\chi})=106.75$.
Although $\chi$ also decays into gluinos,
which might lead to overproduction of LSPs,
they can annihilate before decoupling from the thermal bath.
LSPs produced from the flaton field are negligible
in the case of the wino/higgsino LSP,
which is compatible with the sequestering models 
(see Sec.~\ref{subsec:sequestering}).

We can estimate the dilution factor $\Delta$ as follows:
\begin{eqnarray}
	\Delta&=&\frac{4}{3}\frac{V_0}{2\pi^2/45g_{*s}(T_c)T_c^3T_{\rm RH}^{\chi}} \nonumber \\
	&\simeq&1.3\times10^{12}|\lambda_X|^{-1}\left(\frac{m_{3/2}}{300{\rm\,TeV}}\right)^3
	\left(\frac{300{\rm\,TeV}}{T_c}\right)^3\left(\frac{470{\rm\,GeV}}{T_{\rm RH}^{\chi}}\right).
\end{eqnarray}
One can find that the estimated dilution factor is the same order with the required one in Eq.~(\ref{eq:Delta})
when $T_c\simeq300{\rm\,TeV}$.
Assuming that $m_0\ll T_c\simeq{\cal O}(300){\rm\,TeV}$,
the coefficient $c_T$ should be taken as $c_T\simeq g_{\xi}^{2}\simeq (m_0/T_c)^2$.

In the context of the thermal inflation,
we need to take into account a secondary oscillation of the Polonyi field.
Its potential during the thermal inflation is given by
\begin{eqnarray}
	V&=&\frac{1}{2}m_Z^2z^2+\frac{c'_H}{2}H^2\left(z-z'_0\right)^2 \nonumber \\
	&=&\frac{1}{2}\left(m_Z^2+c'_HH^2\right)
	\left(z-\frac{c'_HH^2}{m_Z^2+c'_HH^2}z'_0\right)^2+\cdots,
\end{eqnarray}
where $z$ denotes the amplitude of $Z$ ($z\equiv|Z|/\sqrt{2}$),
and $z'_0$ represents the local minimum determined by the Hubble induced terms.
$c'_H$ is a coefficient of the Hubble mass term and is of ${\cal O}(1)$.
Note that $z'_0$ is expected to be of the order of the Planck scale.
One can find that
the Polonyi field does not sit at the true minimum 
but at the local minimum determined by $c'_H (H_{\rm th}/m_Z)^2z'_0$
just after the thermal inflation,
where $H_{\rm th}$ represents the Hubble parameter during the thermal inflation
and $H_{\rm th}\ll m_Z$.
Therefore,
the secondary oscillation occurs with the amplitude of $c'_H (H_{\rm th}/m_Z)^2z'_0$.
The abundance of LSP produced from the above contribution
is given by
\begin{eqnarray}
	\frac{\rho_{\rm LSP,sec}}{s_0}&=&2{\rm Br_{\rm SUSY}}
	\frac{m_{\rm LSP}}{m_Z}\left.\frac{\rho_{Z,{\rm sec}}}{s_f}\right|_{H=\Gamma_{\chi}}
	=2{\rm Br_{\rm SUSY}}\frac{m_{\rm LSP}}{m_Z}
	\frac{3T^{\chi}_{\rm RH}}{4V_0}
	\left.\rho_{Z,{\rm sec}}\right|_{H=H_{\rm th}} \nonumber \\
	&\simeq&2.9\times 10^{-19}{\rm\,GeV}c'^2_H
	\left(\frac{\rm Br_{\rm SUSY}}{10^{-3}}\right)
	\left(\frac{T_{\rm RH}^{\chi}}{470{\rm\,GeV}}\right)
	\left(\frac{m_{\rm LSP}}{1{\rm\,TeV}}\right)
	\left(\frac{z'_0}{M_{\rm pl}}\right)^2,
\end{eqnarray}
where we use $m_Z\simeq m_{3/2}$.
Since this is much smaller than the observed DM density,
one can find that the density of LSPs produced from the secondary oscillation
is negligible.

Let us comment on domain walls.
Since the superpotential of $X$ has the $Z_4$ symmetry,
there exists four degenerate minima in the potential for $X$
(see Eq.~(\ref{eq:potential_X})).
The flaton randomly falls into one of them after the thermal inflation,
and then domain walls (DWs) are formed.
The DWs dominate the energy density of the universe,
which leads to a cosmological disaster~\cite{Zeldovich:1974uw}.
A bias for the degenerate minima is needed so that 
DWs collapse before they dominate the universe.
In order to avoid the DW problem,
the following condition should be satisfied:\footnote{Assuming that the energy density of DWs obeys the scaling relation $\rho_{\rm DW}\sim\sigma H$,
DWs dominate the universe when $H_{\rm dom}\sim\sigma/M_{\rm pl}^2$.
On the other hand,
DWs decay due to the bias when $H_{\rm dec}\sim\delta V_{\rm bias}/\sigma$.
The condition is derived by requiring $H_{\rm dec}\gg H_{\rm dom}$.
}
\begin{equation}
	\delta V_{\rm bias}\gg \frac{\sigma^2}{M_{\rm pl}^2},
\end{equation}
where $\delta V_{\rm bias}$ represents the bias of the energy density,
and $\sigma$ represents the tension of DWs.
The necessary bias is so small that it does not change the scenario of the thermal inflation.


\end{document}